\documentclass[12pt]{iopart}
\usepackage{theorem,amssymb,mathrsfs}
\usepackage{iopams}  
\def\operatorname#1{{\rm #1}}
\def\<{\langle}\def\>{\rangle}
\def\Tr{\operatorname{Tr}}\def\:{\hbox{\bf :}}
\renewcommand{\leq}{\leqslant}
\def\map#1{{\mathscr{#1}}}

\def\set#1{{\sf #1}}

\def\dim{\operatorname{dim}}\def\adm{\operatorname{dim}}

\def\sH{\set{H}}

\def\qed{$\,\blacksquare$\par}
\def\eg{e.~g. }\def\ie{i.~e. }

\newtheorem{lemma}{Lemma}
\newtheorem{theorem}{Theorem}
\newtheorem{definition}{Definition}
\newtheorem{corollary}{Corollary}
\newtheorem{postulate}{Postulate}
\def\Proof{\medskip\par\noindent{\bf Proof. }}
\def\trnsfrm#1{\mathscr #1}
\def\tA{\trnsfrm A}\def\tB{\trnsfrm B}\def\tC{\trnsfrm C}
\def\tS{\trnsfrm S}
\def\tI{\trnsfrm I}\def\tU{\trnsfrm U}

\def\AA{\mathbb A}\def\AB{\mathbb B}\def\AL{\mathbb L}
\def\cA{{\underline{\tA}}}\def\cB{\underline{\tB}}
\def\cI{{\underline{\tI}}}

\def\Stset{{\mathfrak S}}
\def\Trnset{{\mathfrak T}}
\def\Cntset{{\mathfrak P}}
\begin{document}

\title[No-signaling, dynamical independence, and the local observability principle]{No-signaling,
  dynamical independence, and the local observability principle}

\author{Giacomo Mauro D'Ariano}

\address{Dipartimento di Fisica ``A. Volta'', via Bassi 6, 27100 Pavia, Italy}
\ead{dariano@unipv.it}
\begin{abstract}
  Within a general operational framework I show that a-causality at a distance of "local actions"
  (the so-called {\em no-signaling}) is a direct consequence of commutativity of local
  transformations, \ie of dynamical independence. On the other hand, the tensor product of Quantum
  Mechanics is not just a consequence of such dynamical independence, but needs in addition the
  Local Observability Principle.
\end{abstract}

\vspace{2pc}
\noindent{\it Keywords}: Axiomatical foundations of Quantum Mechanics\hfill\break
\submitto{\JPA}
\maketitle
\section{Introduction}
The quantum correlations due to entanglement are instantaneous. At first sight this may seem useful
for superluminal communications, \eg in a communication scheme in which Alice, using a singlet state
entangled with Bob's spin, communicates to him a bit value $b=0,1$ by measuring either $\sigma_z$ or
$\sigma_x$, respectively, and then Bob tries to determine if its local spin state is an eigenstate
of $\sigma_z$ or of $\sigma_x$. Such a communication scheme was indeed considered in Ref.
\cite{Herbert}), where a strategy for discriminating Bob's non orthogonal states has been devised
based on cloning states into multiple copies via stimulated emission of radiation. However, the
possibility of cloning quantum states was ruled out in
Refs.~\cite{WoottersZurek,dieks82,ghirardi83,yuen86_clon} (for a history of the no-cloning theorem
see Ref.\cite{peres_clonhist}), where it was shown that perfect cloning is impossible as a
consequence of linearity of quantum mechanical transformations, and, as a consequence of the
no-cloning theorem, it is impossible to discriminate with certainty among non orthogonal
states~\cite{single}.

\par From the point of view of proving {\em no-signaling}, \ie the impossibility of superluminal
communications, the no-cloning argument, however, is incomplete, since it doesn't rule out the
possibility of information transmission by other means, e.~g.  by approximate
cloning~\cite{BuzekHillery,gima,bem,Werner_clon}, or probabilistic cloning~\cite{DuanGuo}. For
example, we know that in some cases we can discriminate perfectly among nonorthogonal
states~\cite{Chefles98}, however, with some probability of inconclusive outcome: couldn't this be
used to achieve a superluminal communication with some probability?  Who guarantees that any quantum
operation performed by Alice and Bob cannot be used to make a superluminal communication using some
entangled state?
\par Since neither no-cloning, nor no-state-discrimination impossibility theorems logically imply
no-signaling, an independent rigorous proof of no-signaling is in order, and, indeed, several
authors \cite{grw,ghi,BuschA,svetlichny,Peres2,superluminal} have analyzed the issue and proved
no-signaling.  The first proofs that non-locality of quantum mechanics cannot lead to superluminal
transmission of information has been given in Refs.~\cite{grw,ghi}, and was then generalized to any
trace-preserving quantum operation in Ref.~\cite{superluminal}. A proof of the local state
invariance for trace-preserving quantum operations has also been given in Ref.~\cite{yuenqbc3}.

The "peaceful coexistence" \cite{peaceful} between quantum non-locality and special relativity has
intrigued many physicists for years, on whether the no-signaling condition plays a more fundamental
role, \eg it could be used as an axiom for deriving Quantum Mechanics itself. In this line of
thought some basic features of Quantum Mechanics have been analyzed, such as no-cloning itself. For
example, the no-signaling constraint has been used to derive upper bounds for the fidelity of
cloning transformations~\cite{Gisin,Hardy,ghosh,Pati}.  Later, however, the existence of a
connection between approximate cloning and the no-signaling has been ruled out \cite{superluminal},
and it has been shown that the no-signaling constraint on its own is not sufficient to prevent a
transformation from surpassing the known optimal cloning bounds. More specifically, in Ref.
\cite{GBS} the possibility of using no-signaling as an axiom of Quantum Mechanics has been
considered again, arguing that, once the Born rule is assumed, the linearity of Quantum Mechanics
can be derived from the no-signaling condition. A big step forward in understanding the axiomatic
role played by no-signaling in Quantum Mechanics has been done in Ref. \cite{popescu}. There it has
been shown that, at the purely statistical level, there exist in principle super-quantum
correlations that violate the quantum bound (such as the Tsirelson's bound \cite{Tsirelson} for the
CHSH correlation \cite{CHSH}) without anyway violating the no-signaling condition. Therefore, it is
possible in principle to have a non-locality that is even stronger than the quantum one, however,
still without violating the no-signaling.

The above considerations and the past research history on no-signaling suggest to seek more precise
logical connections between seemingly related issues such as locality, causality, dynamical
independence, and statistical independence, within a general purely operational framework.  In this
paper I will show that, starting from a very general and comprehensive definition of {\em action} by
an agent in a communication scenario, the no-signaling is a direct consequence of commutativity of
local transformations, \ie of dynamical independence. In the process, I will also give an
alternative very general proof of no-signaling in Quantum Mechanics. On the other hand, I will show
that the tensor product of Quantum Mechanics (which leads to no-signaling) is not just a consequence
of dynamical independence, but needs an additional hypothesis, which is the {\em Local Observability
  Principle} \cite{darianoVax2006}. This plays a crucial operational role in reducing the
experimental complexity for experiments on composite systems, reconciling holism with reductionism
in a non-local theory. For a complete account on the operational framework used in the present and a
related axiomatic derivation of Quantum Mechanics, the reader is addressed to Ref.
\cite{darianoVax2006}.

\section{Operational derivation of no-signaling from dynamical independence}

In a purely operational framework, beyond physical theories, in analyzing a communication
scenario we need precise definitions for {\em action} of a transmitting agent, {\em locality} of
actions, and {\em dynamical independence}. As we will see, the dynamical independence is essentially
synonym of existence of local actions, and locality of actions is synonym of commutativity of
transformations, which in turn leads to no-signaling.
\subsection{Action and state}
\begin{definition}[Action]The {\em action} on a object system (due to an agent producing an interaction of the
  object with an apparatus) leads to an object transformation drawn from a set of possible
  transformations, each one occurring with some probability.
\end{definition}
According to our definition, the action is identified with a set $\AA\equiv\{\tA_j\}$ of possible
transformations $\tA_j$ than can occur on the object system. In an ideal situation the apparatus
signals which transformation actually occurred, and the agent has perfect knowledge of all details
of each transformation.  The agent cannot control which transformation occurs, but he can decide
which action to perform, namely he can choose the set of possible transformations $\AA=\{\tA_j\}$.
For example, in an Alice\&Bob communication scenario Alice encodes the different bit values by
choosing between two actions $\AA=\{\tA_j\}$ and $\AB=\{\tA_j\}$ corresponding to two different sets
of transformations $\{\tA_j\}$ and $=\{\tB_j\}$.  The agent has control on the transformation itself
only in the special case when the transformation $\tA$ is deterministic. In the following, wherever
we consider a nondeterministic transformation $\tA$, we always regard it in the context of an
action, namely assuming that there exists a complementary transformation $\tB$ such that the overall
probability of $\tA$ and $\tB$ is unit.

\begin{definition}[State] A {\bf state} is a probability rule for transformations.
\end{definition}
Therefore, $\omega$ is a state means that $\omega(\tA)$ is a map from the set of all possible
transformations to $[0,1]$ satisfying the completeness condition
$\sum_{\tA_j\in\AA}\omega(\tA_j)=1$. We will take the identical transformation $\tI$ as {\em
  no-action} with $\omega(\tI)=1$. In the following for a given physical system we will denote by
$\Stset$ the set of all possible states and by $\Trnset$ the set of all possible transformations.

\subsection{Dynamics as conditioning}
\paragraph{\bf State conditioning.} When composing two transformations $\tA$ and $\tB$, the probability
$p(\tB|\tA)$ that $\tB$ occurs conditional on the previous occurrence of $\tA$ is given by the Bayes
rule for conditional probabilities $p(\tB|\tA)=\omega(\tB\circ\tA)/\omega(\tA)$.  This sets a new
probability rule corresponding to the notion of {\em conditional state} $\omega_\tA$ which gives the
probability that a transformation $\tB$ occurs knowing that the transformation $\tA$ has occurred on
the physical system in the state $\omega$, namely
$\omega_\tA\doteq\omega(\cdot\circ\tA)/\omega(\tA)$ (in the following we will make extensive use of
the functional notation with the central dot corresponding to a variable transformation). One can
see that the present definition of ``state'' leads to the identification {\em
  state-evolution}$\equiv${\em state-conditioning}, entailing a {\em linear action of
  transformations on states} (apart from normalization) $\tA\omega:=\omega(\cdot\circ\tA)$: this is
the same concept of {\em operation} that we have in Quantum Mechanics. Therefore, in the present
context linearity of evolution is just a consequence of the fact that the evolution of states is
pure state-conditioning: this will include also the deterministic case
$\tU\omega=\omega(\cdot\circ\tU)$ of transformations $\tU$ with $\omega(\tU)=1$ for all states
$\omega$---the analogous of quantum unitary evolutions and channels.

\paragraph{\bf Dynamical and informational equivalence.} From the Bayes conditioning it follows that
we can define two complementary types of equivalences for transformations: the {\em dynamical} and
{\em informational} equivalences. The transformations $\tA_1$ and $\tA_2$ are {\em dynamically
  equivalent} when $\omega_{\tA_1}=\omega_{\tA_2}$ $\forall\omega\in\Stset$, whereas they are {\em
  informationally equivalent} when $\omega(\tA_1)=\omega(\tA_2)$ $\forall\omega\in\Stset$. The two
transformations are then completely equivalent when they are both dynamically and informationally
equivalent, corresponding to the identity $\omega(\tB\circ\tA_1)=\omega(\tB\circ\tA_2)$,
$\forall\omega\in\Stset,\;\forall\tB\in\Trnset$. We call {\bf effect} an informational equivalence
class of transformations (this is the same notion introduced by Ludwig\cite{Ludwig-axI}).  In the
following we will denote effects with the underlined symbols $\cA$, $\cB$, and we will write
$\tA_0\in\cA$ meaning that "the transformation $\tA$ belongs to the equivalence class $\cA$", or
"$\tA_0$ corresponds to the effect $\cA$'', or "$\tA_0$ is informationally equivalent to $\tA$".
Since, by definition one has $\omega(\tA)\equiv\omega(\cA)$, we will legitimately write
$\omega(\cA)$ instead of $\omega(\tA)$. Similarly, one has $\omega_\tA(\tB)\equiv \omega_\tA(\cB)$,
which implies that $\omega(\tB\circ\tA)=\omega(\cB\circ\tA)$, which gives the chaining rule
$\cB\circ\tA\in\underline{\tB\circ\tA}$ corresponding to the "Heisenberg picture" evolution of
transformations acting on effects (notice that in this way transformations act from the right on
effects). Now, by definitions effects are linear functionals over states with range $[0,1]$, and, by
duality, we have a convex structure over effects.  We will denote the convex set of effects by
$\Cntset$.

\subsection{The structure of transformations}
\paragraph{\bf Addition of transformations.} The fact that we necessarily work in the presence of
partial knowledge about both object and apparatus corresponds to the possibility of incomplete
specification of both states and transformations, entailing the convex structure on states and the
addition rule for {\em coexistent transformations}, namely for transformations $\tA_1$ and $\tA_2$
for which $\omega(\tA_1)+\omega(\tA_2)\leq 1,\;\forall\omega\in\Stset$ (\ie transformations that can
in principle occur in the same action). The addition of the two coexistent transformations is the
transformation $\tS=\tA_1+\tA_2$ corresponding to the event $e=\{1,2\}$ in which the apparatus
signals that either $\tA_1$ or $\tA_2$ occurred, but does not specify which one. Such transformation
is specified by the informational and dynamical equivalence classes $\forall\omega\in\Stset$:
$\omega(\tA_1+\tA_2)=\omega(\tA_1)+\omega(\tA_2)$ and $(\tA_1+\tA_2)\omega=\tA_1\omega+\tA_2\omega$.
Clearly the composition "$\circ$" of transformations is distributive with respect to the addition
"$+$". We will also denote as $\tS(\AA):=\sum_{\tA_j\in\AA} \tA_j$ the deterministic transformation
$\tS(\AA)$ corresponding to the sum of all possible transformations $\tA_j$ in $\AA$. We can also
define the multiplication $\lambda\tA$ of a transformation $\tA$ by a scalar $0\leq\lambda\leq 1$ as
the transformation which is dynamically equivalent to $\tA$, but occurs with rescaled probability
$\omega(\lambda\tA)=\lambda\omega(\tA)$. Now, since for every couple of transformation $\tA$ and
$\tB$ the transformations $\lambda\tA$ and $(1-\lambda)\tB$ are coexistent for $0\leq\lambda\leq 1$,
the set of transformations also becomes a convex set. Moreover, since the composition $\tA\circ\tB$
of two transformations $\tA$ and $\tB$ is itself a transformation and there exists the identical
transformation $\tI$ satisfying $\tI\circ\tA=\tA\circ\tI=\tA$ for every transformation $\tA$, the
transformations make a semigroup with identity, \ie a {\em monoid}.  Therefore, the set of physical
transformations is a convex monoid.

It is obvious that we can extend the notions of coexistence, sum and multiplication by a scalar from
transformations to effects via equivalence classes.

\subsection{Dynamical independence and local state}
\par A purely dynamical notion of {\bf independent systems} coincides with the possibility of
performing local actions. More precisely, we define
\begin{definition}[Dynamical independence]\label{d:independence}
  Two physical systems are {\em independent} if on the two systems 1 and 2 we can perform {\em local
    actions} $\AA^{(1)}$ and $\AA^{(2)}$ whose transformations commute each other (\ie
  $\tA^{(1)}\circ\tB^{(2)}=\tB^{(2)}\circ\tA^{(1)},\;\forall \tA^{(1)}\in\AA^{(1)},\,\forall
  \tB^{(2)}\in\AB^{(2)}$).
\end{definition}
Notice that the above definition of independent systems is purely dynamical, in the sense that it
does not contain any statistical requirement, such as the existence of factorized states. Indeed,
the present notion of dynamical independence is so minimal that it can be satisfied not only by the
quantum tensor product, but also by the quantum direct sum. As we will see in the following, it is
the local observability principle of Postulate \ref{p:locobs} which will select the tensor product.
In the following, when dealing with more than one independent system, we will denote local
transformations as ordered strings of transformations as follows
$\tA,\tB,\tC,\ldots:=\tA^{(1)}\circ\tB^{(2)}\circ\tC^{(3)}\circ\ldots$. The notion of independent
systems now entails the notion of {\em local state}---the equivalent of partial trace in Quantum
Mechanics.

\begin{definition}[Local state]
  For two independent systems in a joint state $\Omega$, we define the {\em local state} $\Omega|_1$
  of system 1 as the probability rule $\Omega|_1(\tA)\doteq\Omega(\tA,\tI)$ of the joint state
  $\Omega$ with a local transformation $\tA$ only on the system 1 and with system 2 untouched.
\end{definition}
Clearly, the above notion can be symmetrically defined for system 2, and can be trivially extended
to any number of independent systems, with the local state $\Omega|_n$ of the $n$th system
representing the probability rule of the joint state in which all systems are left untouched apart
from system $n$. 
\section{The no-signaling theorem}
\par We are now in position to prove the general no-signaling theorem.
\begin{theorem}[No-signaling]\label{iacausal}
  Any local action on a system does not affect another independent system. More precisely, any local
  action on a system is equivalent to the identity transformation when viewed from another
  independent system. In equations one has
\begin{equation}
\forall\Omega\in\Stset^{\times 2},\forall\AA,\qquad
\Omega_{\tS(\AA),\tI}|_2=\Omega|_2.
\end{equation}
\end{theorem}
\Proof Since the two systems are dynamically independent, for every two local transformations one
has $\tA^{(1)}\circ\tA^{(2)}=\tA^{(2)}\circ\tA^{(1)}$, which implies that
$\Omega(\tA^{(1)}\circ\tA^{(2)})=\Omega(\cA^{(1)}\circ\tA^{(2)})=\Omega(\tA^{(1)}\circ\cA^{(2)})\equiv
\Omega(\cA^{(1)},\cA^{(2)})$. By definition, for $\tB\in\Trnset$ one has
$\Omega|_2(\tB)=\Omega(\tI,\tB)$, and using the addition rule for transformations
and reminding the definition of $\tS(\AA)$, one has
\begin{equation}
\Omega(\tS(\AA),\tB)=\Omega([\tS(\AA)]_{\inf},\cB)=\Omega(\cI,\cB)=:\Omega|_2(\tB).
\end{equation}
On the other hand, we have
\begin{equation}
\Omega_{\tS(\AA),\tI}|_2(\tB)=\Omega((\tI,\tB)\circ(\tS(\AA),\tI)=
\Omega(\tS(\AA),\tB),
\end{equation}
namely the statement.\qed \bigskip Notice how the no-signaling is a mere consequence of our minimal
notion of dynamical independence in Def. \ref{d:independence}. Notice also the consistency with the
dynamical part of the definition of addition of coexistent transformations, \ie conditioning
\begin{eqnarray}
\Omega_{\tS(\AA),\tI}|_2(\tB)=\Omega_{\tS(\AA),\tI}(\tI,\tB)=
\sum_{\tA_j\in\AA}\Omega_{\tA_j,\tI}(\tI,\tB)\frac{\Omega(\tA_j,\tI)}{
\sum_{\tA_j\in\AA}\Omega(\tA_j,\tI)}\nonumber\\
=\sum_{\tA_j\in\AA}\frac{\Omega(\tA_j,\tB)}{\Omega(\tA_j,\tI)}\frac{\Omega(\tA_j,\tI)}{
\Omega(\tI,\tI)}=\sum_{\tA_j\in\AA}\Omega(\tA_j,\tB)=\Omega(\tI,\tB).
\end{eqnarray}
\begin{corollary}\label{coronosignal} One has the logical equivalence
\begin{equation}\label{eq:coronosignal}
\Omega(\tA,\tI)=1\,\Longleftrightarrow\,\Omega(\tA,\tB)=\Omega(\tI,\tB),\;\forall\tB\in\Trnset.
\end{equation}
\end{corollary}
\Proof The implication from the left to the right is trivial. To prove the reverse implication, just
consider an other transformation $\tA^\#$ to complete an action $\AA=\{\tA,\tA^\#\}$. Now
$0=\Omega(\cA^\#,\cI)=\Omega(\cA^\#,\cB)+\Omega(\cA^\#,\cB^\#)$ which implies that 
$\Omega(\cA^\#,\cB^\#)=\Omega(\cA^\#,\cB)=0$. This implies that $\Omega(\cI,\cB)=
\Omega(\cA,\cB)+\Omega(\cA^\#,\cB)=\Omega(\cA,\cB)$.\qed

\medskip\par Assessing the truth of statement (\ref{eq:coronosignal}) implies no-signaling, since if
$\Omega(\tS(\AA),\tI)=1\,\Longrightarrow\,\Omega(\tS(\AA),\tB)=\Omega(\tI,\tB)$, \ie
$\Omega(\tS(\AA),\tB)=\Omega_2(\tB)$ $\forall\tB\in\Trnset$.

\section{The quantum version of no-signaling theorem}
Since assessing the truth of statement (\ref{eq:coronosignal}) implies the no-signaling, in order to
prove no-signaling in Quantum Mechanics we just need to prove validity of (\ref{eq:coronosignal}) in
the quantum case. For this purpose, we need a simple technical lemma that is reported in
\ref{app:lemma}. We can then prove the quantum version of no-signaling.
\begin{theorem}[Quantum version of Corollary \ref{coronosignal}]\label{l:lsi}
For any positive operator $R\in\sH_A\otimes\sH_B$ 
and any generally trace-decreasing quantum operation
$\map{M}$ which acts locally on $\sH_A$, one has
\begin{equation}
\Tr[\map{M}\otimes\map{I}(R)]=\Tr[R]\quad
\Longleftrightarrow\quad 
\Tr_1[\map{M}\otimes\map{I}(R)]=
\Tr_1[R].
\end{equation}
\end{theorem}
\Proof That the identity $\Tr_1[\map{M}\otimes\map{I}(R)]=
\Tr_1[R]$ implies $\Tr[\map{M}\otimes\map{I}(R)]=\Tr[R]$ is
obvious. The converse implication is not obvious. Therefore,
assume that 
\begin{equation}
\Tr[\map{M}\otimes\map{I}(R)]=\Tr[R].\label{eqtr}
\end{equation}
Invariance of trace under cyclic permutation gives
\begin{equation}
\Tr_1[\map{M}\otimes\map{I}(R)]=\Tr_1[(K\otimes I) R],
\end{equation}
whence, one has
\begin{equation}
\Tr_1[\map{M}\otimes\map{I}(R)]=\Tr_1[R]+\Tr_1\{[(K-I)\otimes I] R\}\equiv\Tr_1[R].
\end{equation}
In fact, due to Eq. (\ref{eqtr}), one has
\begin{equation}
\Tr\{[(I-K)\otimes I] R\}=0,
\end{equation}
but according to Lemma \ref{l:parpos} in \ref{app:lemma}, the operator $\Tr_1\{[(I-K)\otimes I] R\}$
is positive, whence, being trace-less, it must be identically zero.\qed
\section{The tensor product and the local observability principle}
The tensor product realization of dynamically independent systems in Quantum Mechanics does not
follow just from the general definition of dynamical independence. Indeed, Definition
\ref{d:independence} does not exclude the quantum mechanical realization in terms of direct sum,
instead of tensor product (see \ref{app:directsum}).
One way of excluding the direct-sum realization is to consider the existence of states for which
the probability factorizes \eg $\Omega(\tA,\tB)=\omega_1(\tA)\omega_2(\tB)$, however, this would
lead to a definition of independence that is not purely dynamical, but also statistical. A way to
exclude the direct sum in a purely dynamical way is to introduce the following {\em Local
  Observability Principle} 
\begin{postulate}[Local Observability Principle]\label{p:locobs} For every composite system there exist
  informationally complete observables made only of local informationally complete observables.
\end{postulate}
We recall the definition of informationally complete observable.
\begin{definition}[Informationally complete observable] An observable
  $\AL=\{l_i\}$ is informationally complete if each effect can be written as a linear combination of
  elements of $\AL$, namely for each effect $l$ there exist coefficients $c_i(l)$ such that
\begin{equation}
l=\sum_ic_i(l)l_i.
\end{equation}
\end{definition}
We call the informationally complete observable {\em minimal} when its effects are linearly independent.  

As a consequence of the duality between the convex set of states and that of effect, one has the
identity of their affine dimensions $\adm(\Stset)=\adm(\Cntset)-1$ (the missing dimension is due to
the normalization condition for states).

\medskip The Local Observability Principle plays a crucial operational role, since it reduces
enormously the experimental complexity, by guaranteeing that only local (although jointly executed!)
experiments are sufficient to retrieve a complete information of a composite system, including all
correlations between the components. The principle reconciles holism with reductionism in a
non-local theory, in the sense that we can observe a holistic nature in a reductionistic way---\ie
locally.  The principle implies the following identity
\begin{theorem}
The affine dimension of the convex set of states $\Stset_{12}$ of a composed system can be written
in terms of the affine dimensions of the convex sets of states $\Stset_1$ and  $\Stset_2$ of the
component systems as
\begin{equation}\label{admbound1}
\adm(\Stset_{12})=\adm(\Stset_1)\adm(\Stset_2)+\adm(\Stset_1)+\adm(\Stset_2).
\end{equation}
\end{theorem}
\Proof We can first prove that the right side of Eq. (\ref{admbound1}) is an upper bound for the
left side.  Indeed, as we have seen, by duality between $\Stset$ and $\Cntset$ the number of
outcomes of a minimal informationally complete observable is given by
$\adm(\Cntset)=\adm(\Stset)+1$.  Now, consider a global informationally complete measurement made of
two local minimal informationally complete observables measured jointly. It has number of outcomes
$[\adm(\Stset_1)+1][\adm(\Stset_2)+1]$.  However, we are not guaranteed that the joint observable is
itself minimal, whence the right side of Eq. (\ref{admbound1}) is just an upper bound.
\par The opposite bounding can be easily proved by considering that a global informationally
incomplete measurement made of minimal local informationally complete measurements should belong to
the linear span of a minimal global informationally complete measurement.\qed
\par Identity (\ref{admbound1}) is the same that we have in Quantum Mechanics as a consequence of
the tensor product structure. In fact one has $\adm(\Stset)=\dim(\sH)^2-1$, and $\dim(\sH_{12})=
\dim(\sH_1)\dim(\sH_2)$, which gives $\adm(\Stset_{12})+1=[\adm(\Stset_1)+1][\adm(\Stset_2)+1]$.
Therefore, the tensor product is not a consequence of dynamical independence in Def.
\ref{d:independence}, but follows from the local observability principle.\bigskip

\appendix
\section{Technical lemma}\label{app:lemma}
\begin{lemma}\label{l:parpos}
For $A\geqslant 0$ operator on $\sH_A$ and $R\geqslant 0$ operator on $\sH_a\otimes\sH_B$ one 
has 
\begin{equation}
\Tr_1[(A\otimes I)R]\geqslant 0.
\end{equation}
\end{lemma}
\Proof For any vector $|\varphi\>\in\sH_A$ one has $\Tr_1[(|\varphi\>\<\varphi|\otimes I)R]\geqslant
0$ , since for any vector $|\phi\>\in\sH_B$ one has
\begin{equation}
\<\phi|\Tr_1[(|\varphi\>\<\varphi|\otimes I)R]|\phi\>=
(\<\phi|\otimes\<\varphi|)R(|\phi\>\otimes |\varphi\>)\geqslant 0,
\end{equation}
due to positivity of $R$. Then, the statement follows by considering a spectral decomposition of
$A$, namely
\begin{equation}
\Tr_1[(A\otimes I)R]=\sum_n a_n\Tr_1[(|\varphi_n\>\<\varphi_n|\otimes I)R]\geqslant 0.
\end{equation}
\section{The direct-sum dynamical independence}\label{app:directsum}
For a direct-sum pair of systems, a local transformation on system 1 works on a joint state as
$\tA^{(1)}=\tA_+\oplus p_\tA\tI_-$, namely, on a joint state $\Omega$ corresponding to
$\rho_+\oplus\rho_-$ one has 
\begin{equation}
\Omega(\tA,\tI)=\Tr[\tA_+(\rho_+)]+p_\tA\Tr[\rho_-].
\end{equation}
Any couple of local transformations on the two ``systems'' commute, since
\begin{eqnarray}
\tA^{(1)}\circ\tB^{(2)}&=(\tA_+\oplus p_\tA\tI_-)(p_\tB\tI_+\oplus\tB_-)=
p_\tB\tA_+\oplus p_\tA\tB_-\nonumber\\ &=
(p_\tB\tI_+\oplus\tB_-)(\tA_+\oplus p_\tA\tI_-)=\tB^{(2)}\circ\tA^{(1)}.
\end{eqnarray}
The probability rule of a joint state on local transformations is
\begin{equation}\Omega(\tA,\tB)=\Tr[p_\tB\tA_+\oplus p_\tA\tB_-(\rho)]=
p_\tB\Tr[\tA_+(\rho_+)]+p_\tA\Tr[\tB_-(\rho_-)],
\end{equation}
which gives the implication in the statement of Corollary \ref{coronosignal}---\ie implying
no-signaling---since $\Omega(\tA,\tI)=1$ is satisfied only for $p_\tA=1$ and trace-preserving
$\tA_+$, which then implies
$\Omega(\tA,\tB)=p_\tB\Tr[\rho_+]+\Tr[\tB_-(\rho_-)]\equiv\Omega(\tI,\tB)$. Notice how also state
conditioning is consistently defined
\begin{equation}\Omega_{\tA,\tI}(\tB)=\frac{p_\tB\Tr[\tA(\rho_+)]+
p_\tA\Tr[\tB(\rho_-)]}{\Tr[\tA(\rho_+)]+p_\tA\Tr[\rho_-]}.\end{equation}
\section*{Acknowledgments} I acknowledge a very useful discussion with P. Perinotti, who proved
Corollary \ref{coronosignal}. This research has been supported by the Italian Minister of University
and Research (MIUR) under program Prin 2005.

\end{document}